\documentclass[varenna]{cimento}

\usepackage{epsfig}
\usepackage{amsmath}
\usepackage{cite}

\title{Ion-atom hybrid systems}

\author{Stefan Willitsch}

\institute{Department of Chemistry, University of Basel, Klingelbergstrasse 80, 4056 Basel, Switzerland}

\shortauthor{Stefan Willitsch}

\begin{document}

\maketitle

\begin{abstract}
The study of interactions between simultaneously trapped cold ions and atoms has emerged as a new research direction in recent years. The development of ion-atom hybrid experiments has paved the way for investigating elastic, inelastic and reactive collisions between these species at very low temperatures, for exploring new cooling mechanisms of ions by atoms and for implementing new hybrid quantum systems. The present lecture reviews experimental methods, recent results and upcoming developments in this emerging field.
\end{abstract}

\begin{figure}[!b]
\hrulefill \\
\footnotesize{To appear in the Proceedings of the International School of Physics Enrico Fermi, Course 189 "Ion Traps for Tomorrow's Applications".}
\end{figure}

\section{Introduction}

"Hybrid" systems of cold ions and atoms have become the subject of intense study in recent years \cite{haerter13b}. The possibility to trap and cool atomic and molecular ions together with neutral atoms in the same region of space has paved the way for investigating interactions between these species at extremely low energies (corresponding to a few millikelvin and below), for engineering new types of quantum systems and for exploring collisional and chemical processes in a new physical regime. 

Early theoretical studies on the properties of mixed ion-atom systems were reported by C\^ot\'e and coworkers in the early 2000s, focusing on ultracold ion-atom collision dynamics \cite{cote00a}, charge transport \cite{cote00b} and the possible formation of mesoscopic quantum systems consisting of a single ion weakly bound to an ensemble of ultracold atoms \cite{cote02a}. Shortly thereafter, proposals for the implementation of a hybrid ion-atom experiment were published by Smith and co-workers \cite{smith03a, smith05a}. Since then, hybrid traps have been developed in a growing number of laboratories \cite{grier09a, zipkes10a, schmid10a, hall11a, rellergert11a, ravi11a, sivarajah12a, haze13a}. In the present lecture, we review recent work in the field and discuss the technology, properties and applications of hybrid ion-atom systems. 

\section{Trapping of ions, atoms and their combination}
\label{trap}

\begin{figure}[t]
\epsfig{file=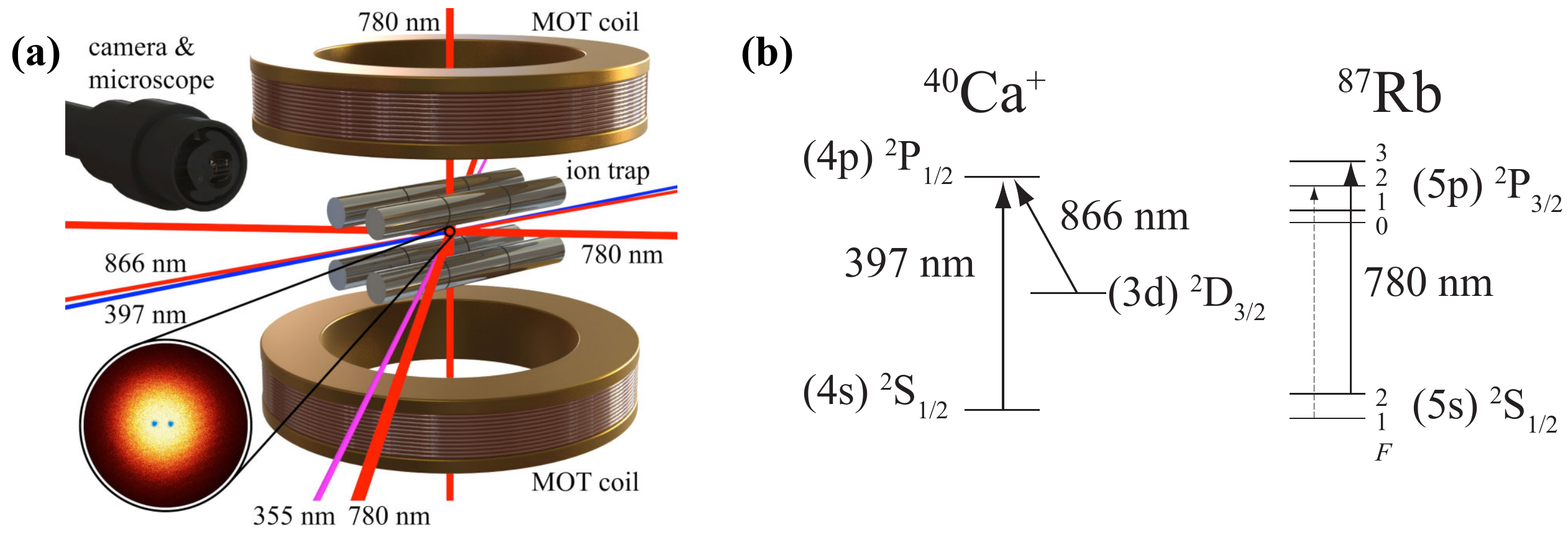, width=\textwidth}
\caption{(a) Schematic of an ion-atom hybrid trap consisting of a magneto-optical trap for atoms superimposed on a linear radiofrequency ion trap for atomic and molecular ions. Inset: false-colour fluorescence image of a Coulomb crystal of two laser-cooled Ca$^+$ ions (blue) embedded in a cloud of ultracold Rb atoms (red-yellow). Reproduced from Ref. \cite{hall13a}. (b) Laser-cooling schemes for $^{40}$Ca$^+$ and $^{87}$Rb.}
\label{setup}
\end{figure}

Mixed systems of cold ions and atoms are produced in combinations of traps for both species. The ion traps conventionally used are radiofrequency (RF) traps which use static and time-varying electric fields to confine charged particles \cite{major05a, willitsch12a}. Periodic voltages applied to the trap electrodes generate an oscillating potential saddle point in the center of the device which enables a dynamic trapping of the ions. Typically, linear RF traps are used in which four electrodes generate a quadrupolar electric potential in the center (see Fig. \ref{setup} (a) and Refs. \cite{drewsen00a, willitsch12a}). Alternatively, surface-electrode ion traps in which all electrodes are situated in a plane have also been employed \cite{grier09a}.

The time varying fields in the trap constantly push and pull the ions back and forth imparting a fast oscillating "micromotion". The micromotion is constantly driven by the RF fields so that its energy is determined by the RF electric field strength at the position of the ion. If the frequency of the micromotion is much larger than the frequency of the thermal ("secular") motion of the ions in the trap, the two types of motion can be adiabatically separated \cite{gerlich92a}. Under these conditions, the time average over the kinetic energy stored in the micromotion gives rise to a time-independent effective trapping potential which governs the secular motion of the ions \cite{major05a, gerlich92a}. 

In typical hybrid trap experiments, atomic ions such as Ca$^+$, Ba$^+$ and Yb$^+$ are used which can efficiently be laser cooled (see Fig. \ref{setup} (b) for the laser cooling scheme used for Ca$^+$). At secular temperatures of a few millikelvin, the laser-cooled ions localise in the trap to form ordered structures termed Coulomb crystals (see inset in Fig. \ref{setup} (a) and Ref. \cite{willitsch12a}). For more complex species such as molecular ions for which laser cooling is not generally feasible, Coulomb crystallisation can nonetheless be achieved by sympathetic cooling through elastic collisions with simultaneously trapped laser-cooled atomic ions \cite{molhave00a, willitsch12a}. The number of ions in Coulomb crystals as well as their secular and micromotion kinetic energies can be determined by comparisons of experimental fluorescence images of the ions with molecular dynamics (MD) simulations as discussed in Refs. \cite{bell09a, hall13a}. The total kinetic energy of the ions in a Coulomb crystal is usually dominated by the micromotion. For a linear quadrupole trap such as the one depicted in Fig. \ref{setup}(a), the time varying fields on the central trap axis vanish. Ions exactly located on the central axis thus exhibit no micromotion and therefore minimal kinetic energies. 

The starting point for the trapping of cold atoms is usually a magneto-optical trap (MOT) in which laser-coolable atoms such as Rb, Yb or Ca are confined and cooled by the combined action of optical and magnetic fields \cite{raab87a} (Fig. \ref{setup} (a)). Typical MOTs consist of two solenoids in anti-Helmholtz configuration generating a quadrupolar magnetic field. Six cooling laser beams in an optical-molasses configuration are used to generate radiation pressure forces on the atoms. Typical temperatures of the atoms in a MOT amount to several hundred $\mu$K. Even lower temperatures down to the nK range at which Bose-Einstein condensation \cite{anderson95a} can occur can be achieved by subsequent evaporative cooling of the atoms after their transfer into a magnetic or optical dipole trap.  

Current hybrid trap experiments rely on the combination of a RF ion trap with a suitable atom trap such as a MOT \cite{grier09a, hall11a, rellergert11a, ravi11a, sivarajah12a}, magnetic or optical dipole trap \cite{zipkes10a, schmid10a, schmid12a}. Fig. \ref{setup}(a) shows the setup implemented at the University of Basel \cite{hall11a, hall13a}. Four segmented cylindrical electrodes forming a linear RF ion trap are sandwiched in between two solenoids generating the magnetic fields of a MOT for $^{87}$Rb atoms. The cooling laser beams for the ions (Ca$^+$ or Ba$^+$) are inserted along the ion-trap axis and intersect with the MOT beams in the center of the combined trap. The fluorescence generated during the laser cooling of both species is imaged onto a camera using a microscope interfaced with the trap. A typical false-colour image showing the superposed fluorescence of both species is shown in the inset of Fig. \ref{setup} (a). Typical densities and temperatures for the ions in this experiment amount to $10^8$~cm$^{-3}$ and 5-10~mK, respectively. The corresponding numbers for the atoms are $10^8$-10$^9$~cm$^{-3}$ and 150-200~$\mu$K, respectively.

\section{Ion-atom interactions: background}

\label{col}

Under the conditions prevalent in hybrid traps, the ions and atoms interact with each other through collisions. The long-range part of the ion-atom interaction potential $V(R)$ is usually expressed in terms of a multipole expansion \cite{krych11a, buckingham67a}:
\begin{equation}
V(R)=\frac{C_3}{R^3}-\frac{C_4}{R^4}+\frac{C_5}{R^5}-...
\end{equation}
where $R$ denotes the inter-particle distance. Terms scaling with $R^{-1}$ and $R^{-2}$ corresponding to the Coulomb and charge-dipole interactions, respectively, which vanish for the interactions of an ion with a neutral atom. The $R^{-3}$ term represents the interaction between the charge of the ion and the permanent electric quadrupole moment of the atom. The $C_3$ coefficient for an atom in the state $\gamma$ with electronic orbital angular momentum $L$ can be expressed as \cite{krych11a, sobelman79a, mies73a}
\begin{equation}
C_3=(-1)^{L-\Lambda}\begin{pmatrix} L & 2 & L \\ -\Lambda & 0 & \Lambda \end{pmatrix} \langle \gamma L \Lambda || Q_2 || \gamma L \Lambda \rangle. \label{cqi}
\end{equation}
In Eq. \ref{cqi}, $\Lambda$ stands for the quantum number of the projection of the electronic angular momentum on the internuclear axis and $\langle \gamma L \Lambda || Q_2 || \gamma L \Lambda \rangle$ stands for the reduced matrix element of the quadrupole moment $Q_2$ \cite{sobelman79a}.

For an ion interacting with an isotropic neutral particle (such as a neutral atom in an $S$ electronic state), the  $C_3$ coefficient vanishes. The then leading $R^{-4}$ term corresponds to the interaction between the charge of the ion and the dipole moment induced in the atom. In this case, the $C_4$ coefficient can be expressed as $C_4=\tfrac{1}{2}\alpha$ where $\alpha$ stands for the isotropic polarisability of the atom. 

Because the majority of present experiments relies on ''heavy'' elements such as Ca, Ba, Yb and Rb, several tens of partial waves contribute to the collisions even at the low temperatures (mK) typically achieved in hybrid traps. Under these conditions, classical models often serve as an adequate starting point to describe the collision dynamics, as long as dedicated quantum effects such as tunneling and scattering resonances \cite{belyaev12a, hall13a} can be neglected. 

When long-range interactions dominate the collisions, centrifugal effects become important. One such example is a "capture" process in which every close-range encounter of the collision partners leads to a chemical reaction \cite{dashevskaya03a}. The centrifugal energy $E_\text{cent}=L^2/2\mu R^2$ is added to the interaction potential yielding an effective, centrifugally corrected potential $V_\text{eff}(R)$. $L$ stands for the collisional angular momentum which is given by $L=\mu vb$ in classical mechanics. Here, $v$ is the collision velocity, $\mu$ is the reduced mass and $b$ the impact parameter, i.e., the shortest distance between the collision partners in the absence of an interaction potential.

\begin{figure}[t]
\begin{center}
\epsfig{file=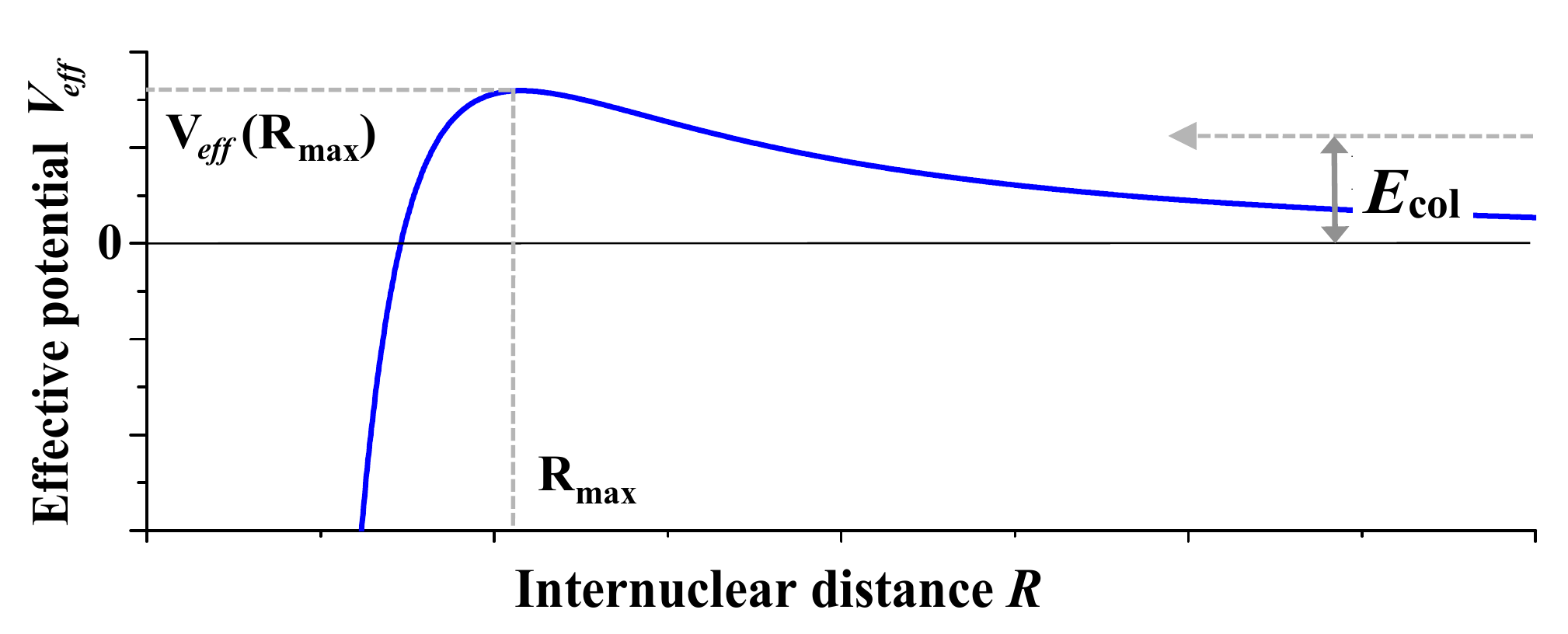, width=0.8\textwidth}
\end{center}
\caption{Schematic representation of the centrifugally corrected long-range interaction potential between an ion and an atom as a function of the internuclear distance $R$. $R_\text{max}$ denotes the position of the maximum of the centrifugal barrier. The collision partners can only approach to close distances if the collision energy $E_\text{col}$ exceeds the height of the barrier.}
\label{pot}
\end{figure}

The centrifugal correction leads to a potential barrier which restricts the range of angular momenta and therefore impact parameters at a given energy (Fig. \ref{pot}). Only collisions which overcome the centrifugal barrier, i.e., which lead to inward-spiralling trajectories and allow the collision partners to approach to close range, lead to a successful reaction. The relevant cross section is given by $\sigma=\pi b_\text{max}^2$ where $b_\text{max}$ is the maximally allowed impact parameter at which the height of the centrifugal barrier does not exceed the collision energy $E_\text{col}$. From this condition, the classical capture cross section for an ion-induced dipole interaction potential (the Langevin cross section) is obtained as $\sigma_L=\pi\sqrt{2\alpha/E_\text{col}}$ \cite{gioumousis58a, levine05a}. The corresponding Langevin rate constant $k_L=\sigma_L v=2\pi\sqrt{\alpha/\mu}$ is independent of the collision energy and thus temperature. 

The complete dynamics including quantum effects, however, can only be captured by quantum scattering calculations \cite{cote00a, bodo08a, tacconi11a, belyaev12a, hall13a, hall13b, zhang09a, zhang09b, rellergert11a, simoni11a, sayfutyarova13a}. Recently, multi-channel quantum-defect theory (MQDT) has been extended to ion-atom collisions \cite{gao08a, idziaszek09a,  idziaszek11a, gao10a, gao11a, gao13a}. MQDT represents an elegant and efficient approach in which the scattering problem can be formulated in terms of only a few system-specific parameters.

\section{Elastic collisions}

Elastic collisions only entail an exchange of kinetic energy. The theory of elastic collisions between cold ions and atoms has been discussed, e.g., by C\^{o}t\'{e} and Dalgarno \cite{cote00a}. In the limit of many partial waves, a useful semiclassical approximation to the elastic scattering cross section $\sigma_\text{el}$ based on the ion-induced dipole long-range interaction is obtained to be \cite{cote00a}
\begin{equation}
\sigma_\text{el}(E_\text{col})=\pi\left(\frac{\mu C_4^2}{\hbar^2} \right)^{1/3}\left(1+\frac{\pi^2}{16} \right)E_\text{col}^{-1/3} \label{sigmael}.
\end{equation}
As the kinetic energy of the ions is typically much larger than that of the ultracold atoms, elastic collisions primarily have two effects.

First, energetic collisions with the ions lead to the ejection of neutral atoms from the shallow atom traps \cite{zipkes10a, schmid10a, haze13a}. As the kinetic energies of the ions are dominated by the micromotion, this effect can be used to probe the micromotion amplitude and better localise the ions on the central trap axis using static control fields \cite{haerter13a}.

Second, the elastic collisions result in a sympathetic cooling of the ions by the ultracold atoms \cite{zipkes10a, schmid10a, ravi12a, sivarajah12a}. Zipkes et al. \cite{zipkes10a} estimate the temperature of a Yb$^+$ ion after sympathetic cooling in a Bose-Einstein condensate of Rb atoms to be 2.1~mK. In a similar experiment, Schmid et al. \cite{schmid10a} report an estimated energy of Ba$^+$ ions in a Rb condensate on the order of $k_\text{B}\times$5~mK. The residual energies of the ions were attributed to uncompensated micromotion. 

The ultimate limit for the ion temperatures which can be reached by sympathetic cooling with ultracold atoms has been the subject of several theoretical studies \cite{zipkes11a, nguyen12a, cetina12a, chen13a}. Cetina et al. predict a micromotion-induced lower limit of the collision energy resulting from the work performed by the trap's RF fields during the atom-ion collision \cite{cetina12a}. Experimental verifications of these predictions remain to be established. The influence of the micromotion could be eliminated by confining the ions in a trap which does not rely on RF fields such as an optical dipole trap \cite{schneider10b}.

\section{Inelastic collisions}

Inelastic collisions entail the change of the internal quantum state of the collisions partners. In particular, they can lead to a relaxation of ions in excited states. These effects have been observed, e.g., in collisions of single state-prepared Yb$^+$ ions with ultracold Rb atoms \cite{ratschbacher12a}.

Inelastic collisions also affect superposition states prepared in an ion implanted into a cloud of ultracold atoms. Ratschbacher et al. \cite{ratschbacher13a} have prepared "qubits" of pairs of specified Zeeman or hyperfine levels in single Yb$^+$ ions and immersed them in a cloud of ultracold, spin-polarised Rb atoms. They observed decay of the excited qubit state resulting from spin relaxation, potentially mediated by spin-orbit interactions. After preparing a coherent superposition between the qubit states, decoherence ($T_2$) times on the order of the collisional time scale were observed, suggesting that spin relaxation is also the dominant mechanism leading to the decoherence of the superposition state. 

While inelastic collisions are undesirable in many cases, they can also have beneficial effects, e.g., for the buffer-gas cooling of the vibrational degrees of freedom of molecular ions (see section \ref{mols}).

\section{Reactive collisions}

Reactive collisions entail a change of the chemical identity of the collision partners. Indeed, in most ion-atom hybrid systems studied so far reactive collisions were shown to play an important role \cite{zipkes10b, schmid10a, hall11a, rellergert11a, ratschbacher12a, hall13a, hall13b}.

\begin{figure}[t]
\begin{center}
\epsfig{file=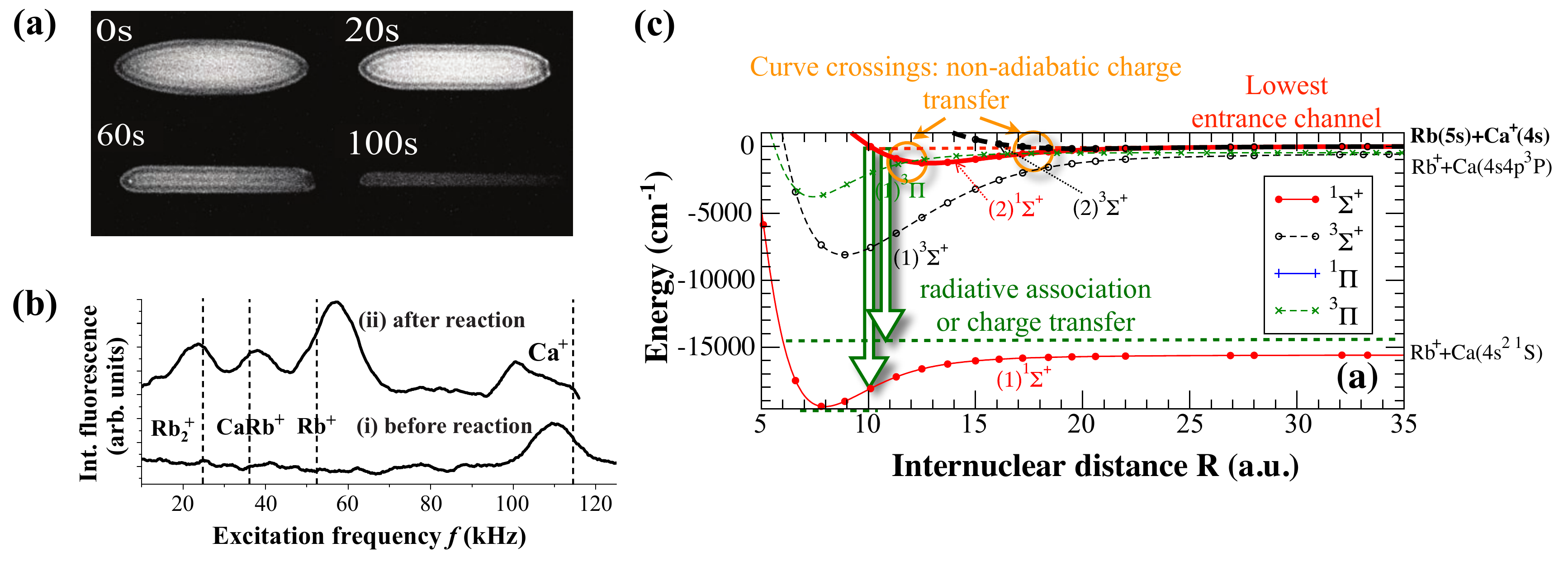, width=\textwidth}
\end{center}
\caption{(a) Fluorescence images of a Ca$^+$ Coulomb crystal immersed in a cloud of ultracold $^{87}$Rb atoms (not shown). Ca$^+$ ions are removed from the crystal during the interaction with the ultracold atoms as a consequence of reactive collisions. Heavier product ions are sympathetically cooled to localise at the extremities of the crystal leading to a characteristic flattening of the Ca$^+$ core. (b) Resonant-excitation mass spectrum of a Coulomb crystal after reaction revealing the presence of Rb$^+$, CaRb$^+$ and Rb$_2^+$ product ions alongside the remaining Ca$^+$ ions. (c) Theoretical potential energy curves of the lowest electronic states of the CaRb$^+$ system to illustrate the important reactive processes non-adiabatic charge transfer, radiative charge transfer and radiative association. See text for details. Figures adapted from Ref. \cite{hall11a}.}
\label{reactions}
\end{figure}

The important reactive processes are illustrated in the prototypical Ca$^+$ + Rb system. Fig. \ref{reactions} (a) shows a large Coulomb crystal of Ca$^+$ ions immersed in a cloud of laser-cooled Rb atoms in a MOT \cite{hall11a, hall13a}. The fluorescence of the Rb atoms has been blocked by a colour filter and is therefore not visible in the images. The volume of the Ca$^+$ crystal shrinks as a function of the time of interaction with the Rb cloud. Simultaneously, the edges of the Ca$^+$ crystal become increasingly flattened, indicating the incorporation of a heavier ion species into the crystal by sympathetic cooling with the remaining Ca$^+$ ion \cite{willitsch12a}. These findings are indicative of a chemical process which removes Ca$^+$ ions from the crystal and supplants them with product ions.

The chemical identity of the product ions can be established using resonant-excitation mass spectrometry \cite{drewsen04a, willitsch12a, hall13a}. The mass spectrum in Fig. \ref{reactions} (b) recorded after exposure of a Ca$^+$ crystal to the ultracold Rb atoms shows four features: one corresponding to the remaining Ca$^+$ ions in the crystal and the three product ions Rb$^+$, CaRb$^+$ and Rb$_2^+$.

The mechanisms leading to the formation of these products are illustrated using the molecular potential energy curves of the CaRb$^+$ system in Fig. \ref{reactions} (c) \cite{hall11a}. Because Rb has a lower ionization energy than Ca, the energetically lowest collision channel (entrance channel) Ca$^+(4s)$+Rb$(5s)$ does not represent the absolute ground state, but corresponds to the excited $(2)~^1\Sigma^+$ molecular state of CaRb$^+$. Through non-adiabatic transitions around curve crossings, the $(1)~^3\Pi$ molecular state can be accessed which correlates asymptotically with the products Rb$^+$+Ca$(4s4p)~^3P$. Exiting the collision on this curve results in a charge transfer (non-radiative or non-adiabatic charge transfer, NRCT) \cite{tacconi11a, belyaev12a}.  

Molecular products can be formed by the emission of a photon from the collision complex to the $(1)~^1\Sigma^+$ electronic ground state of the system. This state asymptotically correlates with the products Rb$^+$+Ca$(4s)^2$. CaRb$^+$ molecular ions are generated by radiative association (RA) through the population of bound vibrational levels of the electronic ground state. Conversely, the population of continuum states on the lowest molecular curve leads to radiative charge transfer (RCT) and the formation of Rb$^+$ ions. Theoretical calculations predict that Franck-Condon factors favour RA over RCT in the lowest collision channel of Ca$^+$+Rb at the low collisions energies achieved in the experiment \cite{hall13a}. The Rb$_2^+$ ions are possibly generated by consecutive reactions of sympathetically cooled CaRb$^+$ product ions with Rb atoms in the MOT.

Note that the particle densities ($n\approx10^8$~cm$^{-3}$) in the MOT-based experiments of Ref. \cite{hall11a, hall13a} appear to be too low to promote the formation of molecular ions through three-body collisions. Such processes have been observed in experiments using significantly denser atom clouds (see below). 

Collisions between Ca$^+$ and Rb in the lowest electronic channel, however, only constitute a minor contribution to the experimentally observed reaction rates. Indeed, the dominant channel was shown to be Ca$^+(4p)$+Rb$(5s)$ which is accessed by the excitation of Ca$^+$ during laser cooling \cite{hall11a, hall13a}. The high reaction rates observed in this channel were explained in terms of the high density of electronic states in this region which provide multiple pathways for NRCT, RCT and RA \cite{hall11a}. 

The reactive mechanisms discussed in the context of Ca$^+$+Rb constitute a model of the chemistry of cold ion-atom hybrid systems. Indeed, similar effects have been observed in Ba$^+$+Rb \cite{hall13b}, Yb$^+$+Ca \cite{rellergert11a}, Ba$^+$+Ca \cite{sullivan12a} and Yb$^+$+Rb \cite{ratschbacher12a} and have been explored in a range of  theoretical studies \cite{knecht10a, krych11a, tacconi11a, rakshit11a, rellergert11a, hall11a, belyaev12a, lamb12a, hall13a, hall13b, sayfutyarova13a}. The importance of the different reactive processes varies across various systems. For instance, for the lowest collision channel the efficiency of RA vs. RCT was calculated to be higher for Ba$^+$+Rb than for Ca$^+$+Rb. This effect was attributed to the more favourable Franck-Condon factors for free-bound transitions in Ba$^+$+Rb caused by the double-minimum structure of the entrance channel potential \cite{hall13b}. 

\begin{figure}[t]
\begin{center}
\epsfig{file=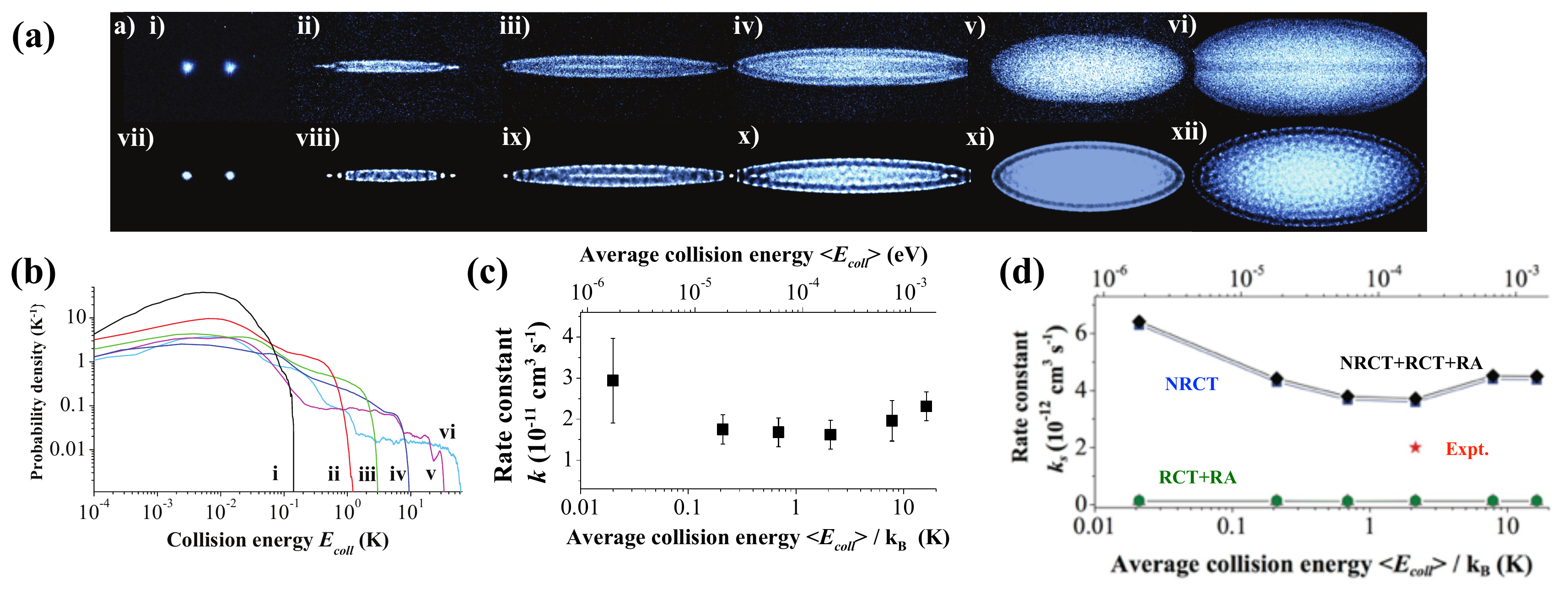, width=\textwidth}
\end{center}
\caption{(a) False-colour fluorescence images of Ca$^+$ Coulomb crystals of various shapes and sizes (i)-(vi) and their molecular dynamics simulations (vii-xii). (b) Collision-energy distributions of the ions in the crystals (i)-(vi) with ultracold Rb atoms. (c) Experimental effective rate constants for Ca$^+$+Rb reactions as a function of the average collision energy. (d) Theoretically predicted rate constants for non-radiative charge transfer (NRCT), radiative charge transfer (RCT) and radiative association (RA) in the lowest collision channel of Ca$^+$+Rb and comparison with experiment. Figures adapted from Ref. \cite{hall13a}.}
\label{ecol}
\end{figure}

An important dynamic characteristic of collisional processes is their collision-energy dependence. In the experiments discussed here, the ion kinetic energies are dominated by the micromotion whose amplitude depends on the position of the ions in the trap (see sec. \ref{trap}). One method to vary the collision-energy distributions in the experiment is by changing the shape and size of the Coulomb crystals \cite{bell09a, hall13a}. Fig. \ref{ecol}(a) (i)-(vi) shows fluorescence images of Ca$^+$ Coulomb crystals of various sizes and shapes together with their MD simulations \cite{hall13a}. Fig. \ref{ecol}(b) shows a double-logarithmic representation of the collision energy distributions of the ions in the crystals (i)-(vi) with cold Rb atoms. The lowest collision energies ($\langle E_\text{col}\rangle/k_\text{B}\approx 20$~mK) have been reached with strings of ions located on the central trap axis (Fig. \ref{ecol} (a)(ii)), whereas large crystals such as the one shown in Fig. \ref{ecol}(a)(vi) exhibit average collision energies as high as $\langle E_\text{col}\rangle/k_\text{B}\approx 20$~K. Fig. \ref{ecol}(c) shows the effective rate constants for reactions of Rb with Ca$^+$ as a function of the average collision energy. The rate constant appears to be practically independent from the collision energy in the interval from 20~mK to 20~K. 

For the reactive processes RCT, RA and NRCT discussed above, the quantum-mechanical reaction cross sections are obtained from the expressions \cite{hall13a, tacconi11a}:
\begin{eqnarray}
\sigma^{RCT}(E_{col})&=p&\frac{8\pi^2}{3c^3}\frac{1}{k^2} \sum_{J=0}^{\infty} \int_{0}^{E_{f}^{max}}\omega^3\left( J | \langle J-1, E_{f}|D(R)|E_{col},J\rangle |^{2}\right. \nonumber \\
&&\left.+(J+1) |\langle J+1, E_{f}|D(R)|E_{col},J\rangle |^{2}\right) dE_{f},
\label{csrct}
\end{eqnarray}
\begin{eqnarray}
\sigma^{RA}(E_{col})&=p&\frac{8\pi^2}{3c^3}\frac{1}{k^2} \sum_{J=0}^{\infty} \sum_{v=0}^{v_{max}}\left( \omega^3 J | \langle J-1,v|D(R)|E_{col},J\rangle |^{2}\right. \nonumber \\
&&\left.+ \omega^3|  \langle J+1, v|D(R)|E_{col},J\rangle |^{2}\right), \label{csra} 
\end{eqnarray}
\begin{eqnarray}
\sigma^{NRCT}=p\frac{\pi\hbar^2}{2\mu E_{col}}\sum_{J=0}^{J_{max}}P_{if}(J,E_{col})(2J+1). \label{nrct}
\end{eqnarray}
Here, $p$ is the statistical weight of the entrance channel, $k$ is the wavenumber of the collision, $E_f$ is the relative energy of the products, $\omega$ is the frequency of the emitted photon, $J$ is the total angular momentum quantum number, $D(R)$ is the electronic transition dipole moment (TDM) between the initial and final state, $\mu$ is the reduced mass and $P_{if}$ is the non-adiabatic transition probability. Fig. \ref{ecol} (d) shows the rate constants computed from the theoretical cross sections \cite{tacconi11a, hall13a} by averaging over the collision-velocity distributions associated with the energy distributions in (b). The red asterisk represents the experimental estimate for the rate constant at a collision energy $\langle E_\text{col}\rangle/k_\text{B}$=2~K.

The observed collision energy dependence can be recovered from the quantum-me\-chanical expressions Eqs. (\ref{csrct})-(\ref{nrct}) by assuming that the TDMs only weakly depend on $J$ and the asymptotic collision energy \cite{hall13a}. The first assumption is justified in the limit of high $J$ applicable to the current experiments ($J>25$). The second assumption holds if the TDMs are localized in a potential well with a depth $D\gg E_\text{col}$ such that the relative velocity at the point of transition is governed by the acceleration through the potential and not the asymptotic collision velocity. This is indeed the case for the lowest collision channel in Ca$^+$+Rb. Under these conditions, it can be shown that the rate constants are independent of the collision energy in agreement with the experimental observations  \cite{hall13a}.

Reactive collision between ions and atoms also give rise to an unsual form of ion cooling mechanism termed "swap" cooling. In Ref. \cite{ravi12a, lee13a}, it was shown that energetic Rb$^+$ ions can undergo charge-exchange collisions with ultracold Rb atoms resulting in the formation of Rb$^+$ ions with a lower kinetic energy. Swap cooling is effective in homonuclear ion-atom systems in which charge exchange leads to  product ions which are chemically identical with the reactant ions. It was shown theoretically in Ref. \cite{ravi12a} that the cooling effects due to the swap mechanism are significant in addition to the sympathetic cooling of ions by elastic collisions. 

Ion-atom hybrid experiments also represent a platform to study three-body collisional processes which become important in dense samples such as Bose-Einstein condensates. Two types of three-body processes have been investigated in this context so far. First, H\"arter et al. \cite{haerter12a} have shown that Rb$^+$ ions implanted in a dense cloud ($n\approx10^{12}$~cm$^{-3}$) of ultracold Rb atoms can serve as a reaction center for three-body recombinations of the form $\mathrm{2~Rb+Rb^+\rightarrow Rb_2 +Rb ^+}$. The bonding energy released in the formation of the Rb$_2$ molecules is converted to kinetic energy of the particles, sending the ions on wide trajectories in the trap before they are re-cooled by the interaction with the ultracold atom cloud. The rate constants determined for the three-body recombinations involving an ion were found to be three orders of magnitude higher than for three colliding neutral atoms.

Second, in a different study H\"arter et al. \cite{haerter13c} have characterised the population distribution of the molecular product states generated in collisions of three neutral ultracold Rb atoms. In these experiments, Rb$_2$ molecules produced in three-body collisions in an ultracold Rb gas were state-selectively ionised and captured in a superimposed ion trap. By scanning the wavelength of the ionization laser, a multiphoton resonance-enhanced photoionization spectrum of the rotational-vibrational quantum states produced in the recombination process was obtained. In this way, insights into the product quantum-state distributions generated in three-body collisions of ultracold atoms could be obtained for the first time.

\begin{figure}[t]
\begin{center}
\epsfig{file=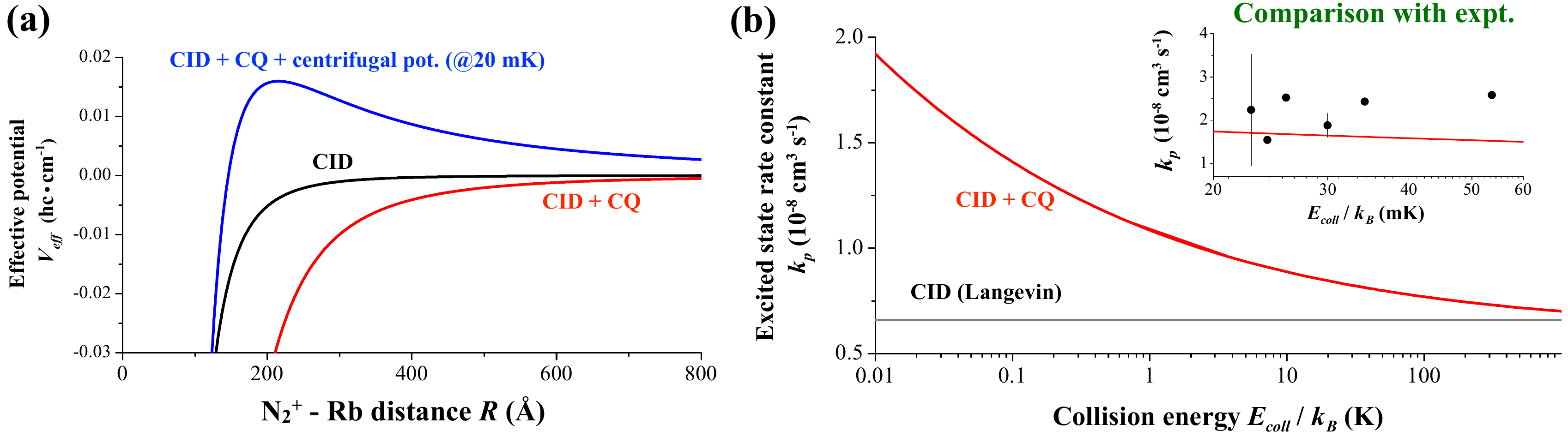, width=\textwidth}
\end{center}
\caption{(a) Charge-induced dipole (CID, black trace) and charge-induced dipole plus charge-permanent quadrupole (CID+CQ, red trace) long-range potentials for the interaction between N$_2^+$ molecular ions and Rb atoms in the $(5p)~^2P_{3/2}$ state. The blue trace represents the centrifugally corrected CID+CQ potential corresponding to the maximum allowed impact parameter at a collision energy $E_\text{col}/k_\text{B}=20$~mK (compare with Fig. \ref{pot}). (b) Predicted classical capture rate constants between N$_2^+$ and Rb~$(5p)~^2P_{3/2}$ assuming long-range potentials including the CID (black trace) and CID+CQ (red trace) interactions. The strong charge-quadrupole interaction leads to larger maximum impact parameters for the collisions and a collision-energy dependent reaction rate constant which is four times larger than the Langevin value at 20~mK. Inset: Comparison with experimental data. Reproduced from Ref. \cite{hall12a}.}
\label{n2}
\end{figure}

\section{Molecular ions in hybrid traps}
\label{mols}

Subtle details of the long-range interaction potential can have significant dynamic effects at low energies. An example is the reaction of sympathetically cooled molecular N$_2^+$ ions with ultracold Rb atoms \cite{hall12a}. The rate for charge-transfer reactions in this system was shown to exhibit a pronounced dependence on the electronic state of Rb ($k<2\times10^{-10}$cm$^{3}$s$^{-1}$ in Rb~$(5s)~^2S_{1/2}$ compared to $k=2.4(13)\times10^{-8}$cm$^{3}$s$^{-1}$ in Rb~$(5p)~^2P_{3/2}$). The high reaction efficiency in the excited state was explained by an electronic near-resonance between the entrance and exit channels \cite{hall12a, vanderkamp94a}. Remarkably, the value of the experimentally determined rate constant in this state is about four times larger than the Langevin rate constant ($k_L=6.6\times10^{-9}$~cm$^3$s$^{-1}$). This large value can be accounted for by the effect of the charge-quadrupole interaction in this channel which is more strongly attractive than the charge-induced dipole interaction which forms the basis for the calculation of the Langevin rate constant (see section \ref{col} and Fig. \ref{n2} (a)). As a consequence, the centrifugal barrier is more strongly suppressed allowing collisions up to higher values of the collisional angular momentum and therefore larger impact parameters. Fig. \ref{n2}(b) shows the theoretical classical rate constants plotted as a function of the collision energy for a charge-induced dipole (CID) as well as a CID plus charge-quadrupole (CQ) interaction potential. Whereas for the CID potential the rate constant is independent from the collision energy, the additional CQ interaction leads to a rate constant scaling with $E_\text{col}^{-1/6}$. The inset in Fig. \ref{n2}(b) shows a magnification of the theoretical capture rate constant for the combined CID and CQ interactions compared with the experimental results. The difference in the rate constants derived with the two potential models is pronounced at the lowest collision energies achieved in the experiment ($E_\text{col}\approx20$~mK). Close to thermal energies at room temperature, however, the difference is only slight and well within the typical experimental error boundaries. 

If the collisions between the molecular ions and atoms are not reactive, they can be used for the sympathetic cooling of the external and internal degrees of freedom of the molecules through elastic and inelastic collisions \cite{hudson09a}. In these cases, the atoms assume the role of an ultracold buffer gas for the molecules. First experiments along these lines were performed by Rellergert et al. \cite{rellergert13a} who recently reported evidence for the cooling of the vibrational motion of BaCl$^+$ molecular ions by collisions with laser-cooled Ca atoms. In their experiments, the population in the vibrational ground state could be increased from 79\% to about 90\% by inelastic collisions with the ultracold atoms.

\section{Conclusions}

Over the past years, ion-atom hybrid experiments have become a vibrant new field. The pioneering works described in the present lecture set the stage for a range of new developments. Exciting new directions include experiments with molecules, new cooling techniques, coherent experiments and the adaption of new types of ion traps which do not suffer from micromotion. 

\section*{Acknowledgments}

This work is supported by the Swiss National Science Foundation grant nr. \linebreak PP00P2\_140834 and the COST Action MP1001 "Ion Traps for Tomorrow's Applications". The author thanks Dr. Ravi Krishnamurthy for critically reading the manuscript.

\bibliographystyle{varenna}

\begin{thebibliography}{10}
\expandafter\ifx\csname url\endcsname\relax\def\url#1{\texttt{#1}}\fi
\expandafter\ifx\csname urlprefix\endcsname\relax\def\urlprefix{URL }\fi

\bibitem{haerter13b}
\NAME{H\"{a}rter A. \atque \mbox{Hecker Denschlag} J.}, \IN{arXiv:1309.5799
  [physics.atom-ph]}{}{2013}{}.

\bibitem{cote00a}
\NAME{C\^{o}t\'{e} R. \atque Dalgarno A.}, \IN{Phys. Rev. A}{62}{2000}{012709}.

\bibitem{cote00b}
\NAME{C\^{o}t\'{e} R.}, \IN{Phys. Rev. Lett.}{85}{2000}{5316}.

\bibitem{cote02a}
\NAME{C\^{o}t\'{e} R., Kharchenko V. \atque Lukin M.~D.}, \IN{Phys. Rev.
  Lett.}{89}{2002}{093001}.

\bibitem{smith03a}
\NAME{Smith W.~W., Babenko E., C\^{o}t\'{e} R. \atque Michels H.~H.}, in
  \TITLE{Coherence and Quantum Optics}, edited by \NAME{Bigelow N., Eberly J.,
  {Stroud Jr.} C.~R. \atque Walmsley I.}, Vol. VIII (Kluwer Academic/Plenum
  Press, New York) 2003, p. 623.

\bibitem{smith05a}
\NAME{Smith W.~W., Makarov O.~P. \atque Lin J.}, \IN{J. Mod.
  Opt.}{52}{2005}{2253}.

\bibitem{grier09a}
\NAME{Grier A.~T., Cetina M., Oru\v{c}evi\'{c} F. \atque Vuleti\'{c} V.},
  \IN{Phys. Rev. Lett.}{102}{2009}{223201}.

\bibitem{zipkes10a}
\NAME{Zipkes C., Palzer S., Sias C. \atque K\"{o}hl M.},
  \IN{Nature}{464}{2010}{388}.

\bibitem{schmid10a}
\NAME{Schmid S., H\"{a}rter A. \atque \mbox{Hecker Denschlag} J.}, \IN{Phys.
  Rev. Lett.}{105}{2010}{133202}.

\bibitem{hall11a}
\NAME{Hall F. H.~J., Aymar M., Bouloufa-Maafa N., Dulieu O. \atque Willitsch
  S.}, \IN{Phys. Rev. Lett.}{107}{2011}{243202}.

\bibitem{rellergert11a}
\NAME{Rellergert W.~G., Sullivan S.~T., Kotochigova S., Petrov A., Chen K.,
  Schowalter S.~J. \atque Hudson E.~R.}, \IN{Phys. Rev.
  Lett.}{107}{2011}{243201}.

\bibitem{ravi11a}
\NAME{Ravi K., Sharma A., Werth G. \atque Rangwala S.~A.}, \IN{Appl. Phys.
  B}{107}{2011}{971}.

\bibitem{sivarajah12a}
\NAME{Sivarajah I., Goodman D.~S., Wells J.~E., Narducci F.~A.,  \atque Smith
  W.~W.}, \IN{Phys. Rev. A}{86}{2012}{063419}.

\bibitem{haze13a}
\NAME{Haze S., Hata S., Fujinaga M. \atque Mukaiyama T.}, \IN{Phys. Rev.
  A}{87}{2013}{052715}.

\bibitem{major05a}
\NAME{Major F.~G., Gheorghe V.~N. \atque Werth G.}, \TITLE{Charged Particle
  Traps} (Springer, Berlin and Heidelberg) 2005.

\bibitem{willitsch12a}
\NAME{Willitsch S.}, \IN{Int. Rev. Phys. Chem.}{31}{2012}{175}.

\bibitem{drewsen00a}
\NAME{Drewsen M. \atque Br{\o}ner A.}, \IN{Phys. Rev.~A}{62}{2000}{045401}.

\bibitem{gerlich92a}
\NAME{Gerlich D.}, \IN{Adv. Chem. Phys.}{82}{1992}{1}.

\bibitem{molhave00a}
\NAME{M{\o}lhave K. \atque Drewsen M.}, \IN{Phys. Rev.~A}{62}{2000}{011401}.

\bibitem{bell09a}
\NAME{Bell M.~T., Gingell A.~D., Oldham J., Softley T.~P. \atque Willitsch S.},
  \IN{Faraday Discuss.}{142}{2009}{73}.

\bibitem{hall13a}
\NAME{Hall F. H.~J., Eberle P., Hegi G., Raoult M., Aymar M., Dulieu O. \atque
  Willitsch S.}, \IN{Mol. Phys.}{111}{2013}{2020}.

\bibitem{raab87a}
\NAME{Raab E.~L., Prentiss M., Cable A., Chu S. \atque Pritchard D.~E.},
  \IN{Phys. Rev. Lett.}{59}{1987}{2631}.

\bibitem{anderson95a}
\NAME{Anderson M.~H., Ensher J.~R., Matthews M.~R., Wieman C.~E. \atque Cornell
  E.~A.}, \IN{Science}{269}{1995}{198}.

\bibitem{schmid12a}
\NAME{Schmid S., H\"{a}rter A., Frisch A., Hoinka S. \atque \mbox{Hecker
  Denschlag} J.}, \IN{Rev. Sci. Instrum.}{83}{2012}{053108}.

\bibitem{krych11a}
\NAME{Krych M., Skomorowski W., Paw{\l}owski F., Moszynski R. \atque Idziaszek
  Z.}, \IN{Phys. Rev. A}{83}{2011}{032723}.

\bibitem{buckingham67a}
\NAME{Buckingham A.~D.}, \IN{Adv. Phys. Chem.}{12}{1967}{107}.

\bibitem{sobelman79a}
\NAME{Sobelman I.~I.}, \TITLE{Atomic Spectra and Radiative Transitions}
  (Springer, Berlin) 1979.

\bibitem{mies73a}
\NAME{Mies F.~H.}, \IN{Phys. Rev. A}{7}{1973}{942}.

\bibitem{belyaev12a}
\NAME{Belyaev A.~K., Yakovleva S.~A., Tacconi M. \atque Gianturco F.~A.},
  \IN{Phys. Rev. A}{85}{2012}{042716}.

\bibitem{dashevskaya03a}
\NAME{Dashevskaya E.~I., Maergoiz A.~I., Troe J., Litvin I. \atque Nikitin
  E.~E.}, \IN{J. Chem. Phys.}{118}{2003}{7313}.

\bibitem{gioumousis58a}
\NAME{Gioumousis G. \atque Stevenson D.~P.}, \IN{J. Chem.
  Phys.}{29}{1958}{294}.

\bibitem{levine05a}
\NAME{Levine R.~D.}, \TITLE{Molecular Reaction Dynamics} (Cambridge University
  Press, Cambridge) 2005.

\bibitem{bodo08a}
\NAME{Bodo E., Zhang P. \atque Dalgarno A.}, \IN{New J.
  Phys.}{10}{2008}{033024}.

\bibitem{tacconi11a}
\NAME{Tacconi M., Gianturco F.~A. \atque Belyaev A.~K.}, \IN{Phys. Chem. Chem.
  Phys.}{13}{2011}{19156}.

\bibitem{hall13b}
\NAME{Hall F. H.~J., Aymar M., Raoult M., Dulieu O. \atque Willitsch S.},
  \IN{Mol. Phys.}{111}{2013}{1683}.

\bibitem{zhang09a}
\NAME{Zhang P., Dalgarno A. \atque C\^{o}t\'{e} R.}, \IN{Phys. Rev.
  A}{80}{2009}{030703}.

\bibitem{zhang09b}
\NAME{Zhang P., Bodo E. \atque Dalgarno A.}, \IN{J. Phys. Chem.
  A}{113}{2009}{15085}.

\bibitem{simoni11a}
\NAME{Simoni A. \atque Launay J.-M.}, \IN{J. Phys. B: At. Mol. Opt.
  Phys.}{44}{2011}{235201}.

\bibitem{sayfutyarova13a}
\NAME{Sayfutyarova E.~R., Buchachenko A.~A., Yakovleva S.~A. \atque Belyaev
  A.~K.}, \IN{Phys. Rev. A}{87}{2013}{052717}.

\bibitem{gao08a}
\NAME{Gao B.}, \IN{Phys. Rev. A}{78}{2008}{012702}.

\bibitem{idziaszek09a}
\NAME{Idziaszek Z., Calarco T., Julienne P.~S. \atque Simoni A.}, \IN{Phys.
  Rev. A}{79}{2009}{010702}.

\bibitem{idziaszek11a}
\NAME{Idziaszek Z., Simoni A., Calarco T. \atque Julienne P.~S.}, \IN{New J.
  Phys.}{13}{2011}{083005}.

\bibitem{gao10a}
\NAME{Gao B.}, \IN{Phys. Rev. Lett.}{104}{2010}{213201}.

\bibitem{gao11a}
\NAME{Gao B.}, \IN{Phys. Rev. A}{83}{2011}{062712}.

\bibitem{gao13a}
\NAME{Gao B.}, \IN{Phys. Rev. A}{88}{2013}{022701}.

\bibitem{haerter13a}
\NAME{H\"{a}rter A., Kr\"{u}kow A., Brunner A. \atque \mbox{Hecker Denschlag}
  J.}, \IN{Appl. Phys. Lett.}{102}{2013}{221115}.

\bibitem{ravi12a}
\NAME{Ravi K., Lee S., Sharma A., Werth G. \atque Rangwala S.~A.}, \IN{Nat.
  Commun.}{3}{2012}{1126}.

\bibitem{zipkes11a}
\NAME{Zipkes C., Ratschbacher L., Sias C. \atque K\"{o}hl M.}, \IN{New J.
  Phys.}{13}{2011}{053020}.

\bibitem{nguyen12a}
\NAME{\mbox{Huy Nguy\^{e}n} L., Kalev A., Barrett M.~D. \atque Englert B.-G.},
  \IN{Phys. Rev. A}{85}{2012}{052718}.

\bibitem{cetina12a}
\NAME{Cetina M., Grier A.~T. \atque Vuleti\'{c} V.}, \IN{Phys. Rev.
  Lett.}{109}{2012}{253201}.

\bibitem{chen13a}
\NAME{Chen K., Sullivan S.~T. \atque Hudson E.~R.}, \IN{arXiv: 1310.5190
  [physics.atom-ph]}{}{2013}{}.

\bibitem{schneider10b}
\NAME{\mbox{Ch.} Schneider, Enderlein M., Huber T. \atque Schaetz T.}, \IN{Nat.
  Photonics}{4}{2010}{772}.

\bibitem{ratschbacher12a}
\NAME{Ratschbacher L., Zipkes C., Sias C. \atque K\"{o}hl M.}, \IN{Nat.
  Phys.}{8}{2012}{649}.

\bibitem{ratschbacher13a}
\NAME{Ratschbacher L., Sias C., Carcagni L., Silver J.~M., Zipkes C. \atque
  K\"{o}hl M.}, \IN{Phys. Rev. Lett.}{110}{2013}{160402}.

\bibitem{zipkes10b}
\NAME{Zipkes C., Palzer S., Ratschbacher L., Sias C. \atque K\"{o}hl M.},
  \IN{Phys. Rev. Lett.}{105}{2010}{133201}.

\bibitem{drewsen04a}
\NAME{Drewsen M., Mortensen A., Martinussen R., Staanum P. \atque Sorensen
  J.~L.}, \IN{Phys. Rev. Lett.}{93}{2004}{243201}.

\bibitem{sullivan12a}
\NAME{Sullivan S.~T., Rellergert W.~G., Kotochigova S. \atque Hudson E.~R.},
  \IN{Phys. Rev. Lett.}{109}{2012}{223002}.

\bibitem{knecht10a}
\NAME{Knecht S., S{\o}rensen L.~K., \mbox{Aa Jensen} H.~J. \atque Marian
  C.~M.}, \IN{J. Phys. B}{43}{2010}{055101}.

\bibitem{rakshit11a}
\NAME{Rakshit A. \atque Deb B.}, \IN{Phys. Rev. A}{83}{2011}{022703}.

\bibitem{lamb12a}
\NAME{Lamb H. D.~L., McCann F.~F., McLaughlin B.~M., Goold J., Wells N. \atque
  Lane I.}, \IN{Phys. Rev. A}{86}{2012}{022716}.

\bibitem{lee13a}
\NAME{Lee S., Ravi K. \atque Rangwala S.~A.}, \IN{Phys. Rev.
  A}{87}{2013}{052701}.

\bibitem{haerter12a}
\NAME{H\"{a}rter A., Kr\"{u}kow A., Brunner A., Schnitzler W., Schmid S. \atque
  \mbox{Hecker Denschlag} J.}, \IN{Phys. Rev. Lett.}{109}{2012}{123201}.

\bibitem{haerter13c}
\NAME{H\"{a}rter A., Kr\"{u}kow A., Dei{$\ss$} M., Drews B., Tiemann E. \atque
  \mbox{Hecker Denschlag} J.}, \IN{Nat. Phys.}{9}{2013}{512}.

\bibitem{hall12a}
\NAME{Hall F. H.~J. \atque Willitsch S.}, \IN{Phys. Rev.
  Lett.}{109}{2012}{233202}.

\bibitem{vanderkamp94a}
\NAME{\mbox{van der Kamp} A.~B., Cosby P.~C. \atque \mbox{van der Zande}
  W.~J.}, \IN{Chem. Phys.}{184}{1994}{319}.

\bibitem{hudson09a}
\NAME{Hudson E.~R.}, \IN{Phys. Rev. A}{79}{2009}{032716}.

\bibitem{rellergert13a}
\NAME{Rellergert W.~G., Sullivan S.~T., Schowalter S.~J., Kotochigova S., Chen
  K. \atque Hudson E.~R.}, \IN{Nature}{495}{2013}{490}.

\end{thebibliography}

\end{document}